# Equation of state of surface-adsorbing colloids

Robert D. Groot and Simeon D. Stoyanov
Unilever Research Vlaardingen,
P.O. Box 114, 3130 AC Vlaardingen, The Netherlands

We have developed a simulation model to describe particle adsorption to and desorption from liquid interfaces. Using this model we formulate a closed interfacial equation of state for repulsive elastic spheres. The effect of a long-range attractive interaction is introduced by perturbation theory, and the effect a short-range attraction is studied using direct simulation. Based on our model predictions we conclude that for polymeric particles the surface pressure *cannot* be modelled directly by inert particles that interact via some effective potential. Internal degrees of freedom within gel particles are all-important. Consequently, the surface pressure of a fully packed layer is not proportional to $kT/d^2$, where $d$ is the particle diameter; but proportional to $kT/d_m^2$, where $d_m$ is the size of the *molecular* units that make up the particle. This increases the surface pressure and modulus by some four orders of magnitude. For short range interaction we study the dynamic behaviour, and find fractal-like structures at low concentrations. At intermediate coverage an irregular structure is formed that resembles a spinodal system. This system freezes, which arrests the spinodal structure. At high surface coverage the simulations show poly-crystalline domains. For dilute systems, the strength of the surface layers is determined by simulated compression and expansion. We find a power law for the surface pressure ($\Pi \sim \Gamma^{10\pm0.5}$), which is related to the (fractal) structure of the adsorbed network. The power law is consistent with surface percolation.

## 1. Introduction

The stability of aerated products, which are usually stabilised by proteins, surfactants or fat globules, is to a large extent limited by the process of Ostwald ripening, that leads to coarsening of foam and eventually complete loss of air from a product.[1,2,3] This problem can be partly avoided by gelling the continuous phase, but a drawback arises from the limited variety of textures that can be engineered in this way.

In the last ten years attention has been drawn again toward the use of colloidal particles (10-1000 nm) to stabilize emulsions or foams.[4] Stabilisation of foams by particles is beneficial for a number on reasons: First of all, particles having a contact angle close to 90° have an adsorption energy that is much higher than that of proteins or surfactants.[5] Furthermore, because of their relatively large size they form a steric layer around bubbles, preventing coalescence. Moreover, they form highly rigid and elastic layers around bubbles preventing Ostwald Ripening. In addition to contact angle, particle shape[6] and particle-particle interactions play crucial roles in the stability of a foam as pointed out by Binks.[5,7]

To engineer particle-stabilised foams we need to understand the physical properties of the air-water interface with adsorbed particles quantitatively from a valid thermodynamic model. The starting point for such models is the surface isotherm, a 2D equation of state that relates spreading pressure $\Pi$ to the adsorbed amount $\Gamma$. Traditionally, a number of equations is used. For the application of surfactants at interfaces, an overview is summarized by Kralchevsky *et al.*[8] The simplest equation of state is the Henry equation [$\Pi = \Gamma kT$] which is the 2D analogue of an ideal gas equation. The Langmuir [$\Pi = -\Gamma_0 kT \ln(1-\Gamma/\Gamma_0)$] and Volmer [$\Pi = \Gamma kT/(1-\Gamma/\Gamma_0)$] equations introduce corrections for the excluded volume, and in the Frumkin and Van der Waals equations a $\Gamma^2$ cohesive pressure term is added to the Langmuir and Volmer equations respectively.

Frequently used models to describe particle adsorption are Frumkin and Van der Waals equations[4] for particles smaller than 10 nm. For larger particles, variations of these have been suggested by Fainerman *et al.*[9,10] In these models particles at the interface are considered to be inert hard spheres with a soft, long-range attraction. For a quantitative description it is essential to describe the hard sphere pressure correctly. However, the Van der Waals equation is known to describe the hard core effect rather poorly. Consequently, the excluded volume may be wrong by a factor two or more if it is used to fit experimental data. A systematic approach to include the hard core effects was discussed by Rusanov.[11] A second problem is that the surface pressure is generally independent of particle size for a given area fraction, whereas the above equations predict a strong dependence on particle size. Fainerman *et al*[10] explain this by assuming a surface layer of excess solvent that generates osmotic pressure. The solvent molecular area in this model, however, depends upon the choice of a dividing surface which leads to some ambiguity, especially when particles are present at the interface.

In attempt to overcome this, we explore an alternative explanation based on computer modeling of surface-adsorbing spherical particles. The results provide insight in the effects of the range and strength of the interaction potential, which leads to design rules for foams, stabilised by particles. We introduce the model and simulation methods in the next section; then we describe the theory and simulations of adsorbed elastic spheres and present predictions for elastic spheres with long-range attraction. Finally we discuss simulation results of elastic spheres with a short-range attractive interaction, where we study the transient dynamics of the system and its response to compression and expansion. Although the major focus of this work is on air-water interfaces our approach is equally applicable to the oil-water interface, and therefore to emulsion systems, proved that appropriate inter-particle interaction potentials are used.

## 2. Simulation model and methods

### 2.1 Simulation model

This section gives a brief summary of the simulation model. A more detailed account is given elsewhere.[12] The simplest model to represent colloidal particles, is to use elastic spheres with short-range attraction. For elastic spheres, the interaction force remains finite, hence the pressure can be found from the virial theorem. Furthermore, the phase diagram of sticky hard spheres (hard spheres with an attractive force of vanishing range) has no solid phase. In reality spheres with an attractive interaction over a finite range do have a crystalline solid phase.[12] Therefore, the sticky elastic sphere model is more realistic than the adhesive hard sphere model. The repulsive force in this model is taken as

$$F^{Rep} = a\left(1 - \frac{r}{d}\right) \quad (r < d) \qquad 1$$

Here, $r$ is the distance between particle centres and $d$ is the particle diameter. The pre-factor $a$ can be related to the linear elastic modulus and the particle size,[12] by $a = \pi E d^2/6$, where $E$ is the elastic modulus. At separation distances further away than a particle diameter, the force should become attractive because particles at contact form hydrogen bonds, or have a strong hydrophobic interaction. A simple form to represent this is to assume a parabola force,

$$F^{Att} = \frac{4\varepsilon}{\delta^2}(1 - r/d)(1 + \delta - r/d) \qquad 2$$

The parameter $\delta$ is the range of the attractive interaction relative to the particle diameter, and $\varepsilon$ is the force minimum.

One might be tempted to take an extremely small value for $\delta$, e.g. because the range of a hydrogen bond is very small. However when a bond is formed, the distance between particle centres may still vary, even if the surfaces are glued together with no intervening space, because the particles are flexible. Therefore, the range of the attractive interaction is actually set by the modulus of the particles, or by the salt strength in case of charged particles, and not by the range of the hydrogen bond.

If we require that the force derivative is continuous in $r = d$, the force amplitude and range are related by $\varepsilon = \frac{1}{4} a\delta$. The force range thus follows from the linear elastic modulus $E$ of the particle and adhesion energy $G$ as[12]

$$\delta = \sqrt{\frac{6G}{EV}} \qquad 3$$

where $V = \pi d^3/6$ is the particle volume. Likewise, the force amplitude follows as

$$\varepsilon = \frac{3}{2d}\sqrt{EVG/6} \qquad 4$$

As a thermostat, pair-wise noise and friction are added between neighbours within a cut-off distance $r_c$ as in standard dissipative particle dynamics.[13,14] The dissipative and random forces between neighbouring particles $i$ and $j$ are given by $\mathbf{F}_{ij}^D = -\gamma(1-r_{ij}/r_c)^2(\mathbf{v}_{ij}\cdot\mathbf{e}_{ij})\mathbf{e}_{ij}$ and $\mathbf{F}_{ij}^R = \sigma(1-r_{ij}/r_c)\theta_{ij}(t)\mathbf{e}_{ij}/\sqrt{\delta t}$, where $\mathbf{e}_{ij}$ is a unit vector pointing from $i$ to $j$, and $\theta_{ij}(t)$ is a random noise variable of variance 1. The noise and friction amplitudes are related via $\frac{1}{2}m\sigma^2 = \gamma kT$. The use of pair-wise noise and friction ensures that momentum is conserved. The hydrodynamic friction between particles of unequal size is a non-trivial extension, which was worked out by Farr et al.[15] Here we restrict ourselves to monodisperse particles.

It should be noted that our unit of time throughout the paper is the natural unit of time, which follows from taking thermal energy as unit of energy ($kT = 1$), and choosing the particle diameter as unit of length ($d = 1$). Hence, the unit of time is $\tau = d(m/kT)^{1/2}$. At high volume fractions of particles this time unit may be gauged to physical time by matching the collective diffusion constant of the colloid to its physical value, see e.g. Groot and Rabone.[16] However, in the present simulations, solvent is not taken into account explicitly, therefore the diffusion behaviour at low concentrations is overestimated. Thus, the time scale mentioned is only a qualitative measure of physical time.

### 2.2 Simulating adsorption equilibrium

Because binding particles to the air-water interface locks the particle motions to a length scale much smaller than their size, we want to model a process where particles are either free or bound into an infinitely thin layer. This can be realised by a surface association equilibrium as follows.

When a particle hits the surface it sticks to it with a given probability. Every subsequent time-step the adsorbed particles are desorbed with probability $p$. The adsorption equilibrium now results from two rate processes: adsorption due to particles hitting the wall and desorption due to a probabilistic process. At infinite dilution $\rho$ of the adsorbing species, the adsorption process is characterised by the adsorption constant $K_{wall}$

$$\Gamma = K_{wall}\rho + O(\rho^2) \qquad 5$$

where $\Gamma$ is the number of adsorbed particles per unit of surface area, and $\rho$ is the concentration of adsorbing particles in the bulk.

The surface association constant $K_{wall}$ is the quantity of physical relevance, rather than the adsorption probability. The (ideal) free energy change on adsorption is just $-\ln(K_{wall})$ for every particle that adsorbs on the surface. There is a close analogy between adsorption to a surface and association in bulk. The association constant in bulk is related to the second virial coefficient.[17,18,19] By analogy, $K_{wall}$ can be found as

$$K_{wall} = \int_0^\infty (e^{-u(x)/kT} - 1)dx \qquad 6$$

where $u(x)$ is the interaction energy between the air phase and particle in solution.

Since the adsorption process depends on the chosen time step $\delta t$, the desorption probability $p$ used in the simulation must depend on the time-step in a particular way to ensure that the (physical) adsorption constant $K_{wall}$ does not depend on the time step. We first consider the adsorption process. This is determined by the number of particles that arrive at the surface during one time step $\delta t$, and is given by $N_{coll} = \rho \delta t \sqrt{kT/2\pi m}$. When a particle hits the wall, it will remain there with probability $\omega(1-p)$. Thus, the adsorption rate is found as $K_{ads} = \omega(1-p) N_{coll} = \omega(1-p)\rho \delta t \sqrt{kT/2\pi m}$, where $0<\omega<1$ is a factor that differentiates between diffusion limited adsorption ($\omega \sim 1$) and reaction limited adsorption ($\omega \sim 0$). One may interpret $-kT \ln(\omega)$ as the energy barrier to adsorption.

The desorption rate (per time step) is given by $K_{des} = p\Gamma$. At equilibrium an equal number of particles adsorb and desorb, hence $K_{des} = K_{ads}$. The adsorption constant is thus found as $K = \Gamma/\rho = (1-p)/p \; \omega \delta t \sqrt{kT/2\pi m}$, from which we find the desorption probability as

$$p = \frac{\omega \delta t \sqrt{kT/2\pi m}}{K_{wall} + \omega \delta t \sqrt{kT/2\pi m}} \qquad 7$$

The reason that a factor $(1-p)$ is included in the adsorption probability is to ensure that $0 < p < 1$, even for very small values of $K_{wall}$. If this factor is left out $p$ cannot be properly normalised. The interpretation of the factor $(1-p)$ is that particles adsorb in one step (with probability 1) and desorb again in the same step (with probability $p$). This ensures adsorption equilibrium in that step. The simulation method is checked by adsorbing ideal gas particles in a small box; see Footnote ‡.

## 3. Elastic spheres with long-range attraction

### 3.1 Simulation of repulsive spheres

To model the adsorption of particles to an interface quantitatively, we first need to specify the interaction force between particles and surface. Electrostatic dipoles cause the particles to dip into the interface and deform the surface. This may induce a long-range (capillary) force between particles at or near a surface.[20,21] This lateral interaction typically behaves as

$$U^{cap} \sim r^{-4} \qquad 8$$

Moreover, irregular particle shapes may lead to undulations of the meniscus, which induces a similar long-range potential.[22] As a further complication, these capillary forces are non-additive. Moreover, in the case of oil-water interfaces the nature of the polar phase is very important.[23] Since these effects are quite subtle and complicated, direct simulation is very challenging. To attack the problem in an alternative way is to study the effect of various short-range and long-range interaction forces, and then observe the consequences of such forces.

Only few attempts have been made to simulate the 2D equation of state directly if a long range interaction is present. This shows a liquid phase and two solid phases.[24] The solid-solid transition has been observed before, and is identified as a transition between hexagonal and hexatic.[25] For the attractive interaction we make a distinction between a long-range ($\delta > 1$) and short-range ($\delta \ll 1$) interaction. If the interaction range is similar to or larger than the particle size, we may use perturbation theory, and take the repulsive disk as a reference. We will study this here and use direct simulation to study the second case in the next section.

Our first objective is thus to obtain the equation of state of repulsive elastic disks. As a starting point we use the equation of state for hard disks by Henderson,[26]

$$\frac{\Delta \Pi^{HD}}{\Gamma kT} = Z(y) = \frac{2y}{1-y} + \frac{9}{8}\frac{y^2}{(1-y)^2} \qquad 9$$

where $\Gamma = N/A$ is the surface density, $y = \pi \Gamma d^2/4$ is the surface fraction of the hard disks, and $\Delta \Pi^{HD}$ is the hard disk excess pressure. More recent simulations of the hard disk equation of state are reported by Santos et al[27] and by Kofala and Rottner.[28] Though slightly more accurate at higher densities, these expressions are much more complicated so we will use the Henderson equation which is accurate enough for our purpose.

In general, the surface pressure follows from the free energy per particle $f_n$ as $\Pi = \Gamma^2 \partial f_n/\partial \Gamma$, and the corresponding chemical potential is given by $\mu = \partial f_a/\partial \Gamma = f_n + \Pi/\Gamma$, where $f_a$ is the free energy per unit of area. Once we have determined the pressure as function of surface coverage $\Gamma$, the free energy and chemical potential follow consistently by integration of $\Pi/\Gamma^2$. The hard disk chemical potential is thus found as[29]

$$\frac{\Delta m^{HD}}{kT} = G(y) = -\frac{7}{8}\ln(1-y) + \frac{25}{8}\frac{y}{(1-y)} + \frac{9}{8}\frac{y^2}{(1-y)^2} \qquad 10$$

For completeness, the excess free energy per particle is also given here as $\Delta f_n/kT = -7/8 \ln(1-y) + 9/8 \; y/(1-y)$. To find the *elastic* disk equation of state, the pressure is determined by simulation for a range of densities and repulsion parameters, and the *excess* pressure is fitted to the scaling function

$$\frac{\Delta \Pi}{\Gamma kT} = \frac{b}{l} Z(l y) \qquad 11$$

where $Z(y)$ is the hard disk excess compressibility factor given in Eq 9. For the interaction force given in Eq 1, parameter $b$ in Eq 11 follows analytically from the ratio of the second virial coefficients of the elastic disk fluid and the hard disk fluid. Using the hard disk diameter as unit of length ($r^* = r/d$), we obtain

$$b = 2\int_0^1 \left[1 - \exp\left(-\tfrac{1}{2}ad(1-r^*)^2/kT\right)\right] r^* dr^*$$
$$= 1 - \sqrt{\frac{2pkT}{ad}} \, \mathrm{erf}(\sqrt{ad/2kT}) + \frac{2kT}{ad}\left[1 - \exp(-ad/2kT)\right] \qquad 12$$

Simulations were run over $2 \times 10^4$ steps of $\delta t = 0.005$, for repulsion parameters $ad/kT$ = 10, 30, 300, 1000, 3000 and $10^4$, after some $10^4$ steps of equilibration. The data is fitted to the functional form given in Eq 11, where λ is the only fit parameter. One may view √λ as the effective hard disk diameter of the elastic disks. The obtained fit parameters are given in Table I. Note that the values for $b$ are calculated from Eq 12, which is exact. The simulated excess pressure and the fits are shown in Figure 1a. Figure 1b shows the scaled simulation data that fall on a single master curve, confirming that the data collapse is really excellent.

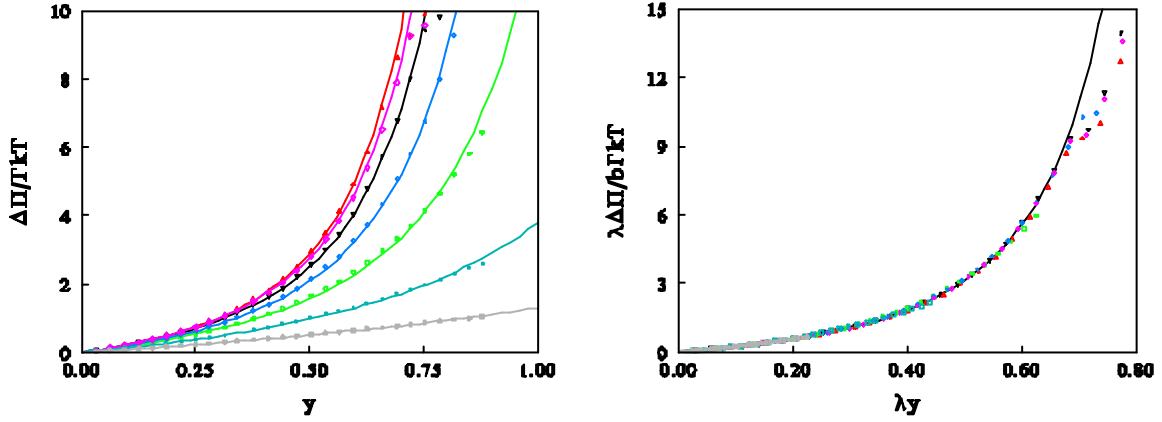

**Figure 1a** (left): Excess pressure and fit functions for the simulation parameters given in Table I. **b** (right): simulation data scaled to one master graph. The data points that deviate from the scaling curve correspond to the solid state.

To obtain a closed expression for the elastic disk equation of state, a smooth fit function was obtained for the scaling parameter λ as function of the repulsion parameter $a$:

$$\lambda \approx 1 - \frac{2.91}{\sqrt{ad/kT}}\left[\exp(-\frac{ad}{3\cdot 10^4 kT}) - \exp(-\frac{\sqrt{ad/kT}}{1.91})\right] \quad \mathbf{13}$$
$$\approx 1.28\,b - 0.27$$

The former expression gives the density scaling factor to within a maximum error of 0.3%, and a mean error of 0.1%. This is accurate enough for a realistic representation of the data, as shown in Figure 1a where the fit curves are based on this interpolation function, and not on the actual fit parameters given in Table I. It is also noted that λ and $b$ are linearly related to within an error of 0.004. This proportionality only fails at very low and very high repulsion parameters, but it does indicate that λ should follow a similar functional form as $b$. Since $b$ is known analytically, we may use this relation to find a suitable functional form of λ for other repulsion forces than linear springs. These are briefly discussed below.

For any elastic disk fluid, the pressure and chemical potential can now be obtained simply by rescaling the density to a lower value, and taking the pressure and chemical potential at that density up to a factor. Thus,

$$\Pi(\Gamma,a) = \Gamma kT + b\Gamma kT\, Z(\lambda y)/\lambda$$
$$\mu(\Gamma,a) = kT\ln(\Gamma) + bkT\, G(\lambda y)/\lambda \quad \mathbf{14}$$

where $y = \pi\Gamma d^2/4$ is the area fraction, and $Z(y)$, $G(y)$, $b(a)$ and $\lambda(a)$ are given in Eqs 9, 10, 12, and 13.

**Table I** Simulated Interaction Parameters and Fit Results[a]

| $ad/kT$ | $b$ | λ |
|---|---|---|
| 10 | 0.407228 | 0.2565±0.002 |
| 30 | 0.609021 | 0.4994±0.003 |
| 100 | 0.769337 | 0.7116±0.002 |
| 300 | 0.861946 | 0.8332±0.001 |
| $10^3$ | 0.922733 | 0.9116±0.0005 |
| $3\times 10^3$ | 0.954902 | 0.9488±0.0004 |
| $10^4$ | 0.975134 | 0.9814±0.0007 |

[a] The values given for $b$ follow from Eq 12 and are exact. The column marked λ gives the fit parameters to simulated equation of state, Eq 11.

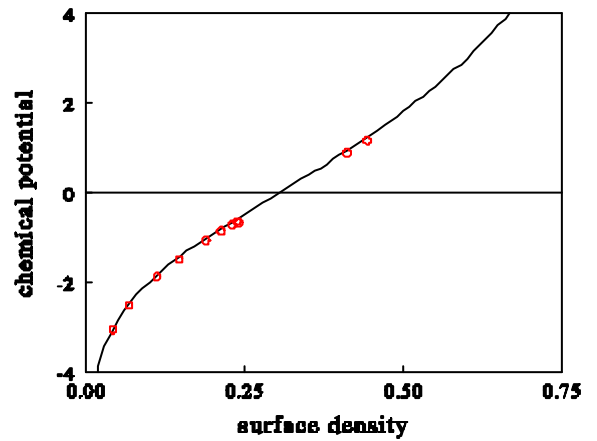

**Figure 2** Chemical potential measured via adsorption equilibrium (red dots) and from elastic disk equation of state, Eq 14.

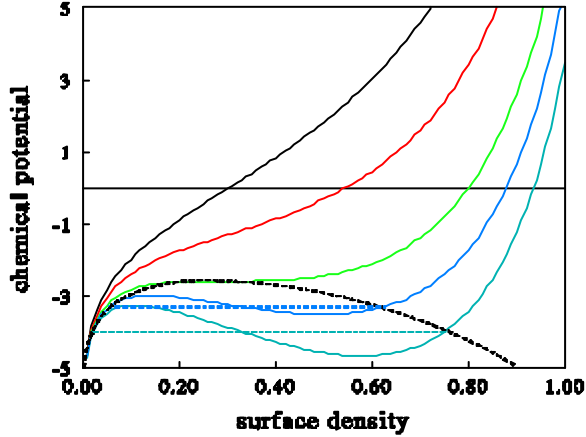

**Figure 3** Chemical potential of elastic disks with long range interaction calculated for $ad/kT$ = 1000, with $b_2$ = 0, 0.5 $b_2^c$, $b_2^c$, 1.25 $b_2^c$, and 1.5 $b_2^c$. The critical point for this repulsion is found as $b_2^c$ = 5.432.

To determine the chemical potential directly as a check, 200 particles of repulsion parameter $a$ = 1000 were simulated in a box of size 5×28×28. Another 200 ideal particles were added to serve as a thermostat. In this simulation the noise and friction weight function was cut off at radius $r_c$ = 2. The DPD thermostat was used with friction factor $\gamma$ = 1, and the elastic spheres were allowed to adsorb on the left wall with adsorption constant $K_{wall}$ = 1, 2, 5, 10, 25, 50, 100, 150 and 200. Another two simulations were done for $K_{wall}$ = 100 and 200, using 400 elastic spheres and 200 ideal particles, which allows investigation of higher adsorbed amounts.

For these systems the equilibrium bulk density was determined by averaging the bulk density profile between $x$ = 2 and $x$ = 5. Since the typical bulk volume fractions are low (0.1% to 2%), we can safely use the Carnahan-Starling equation of state of hard spheres to calculate the chemical potential in the bulk.[29] Since the bulk system and the surface system are in equilibrium via the adsorption and desorption processes, the chemical potential in the surface layer must equal that of the bulk, up to an additional adsorption term that we find from the ideal behaviour $\ln(\Gamma) = \ln(\rho)+\ln(K_{wall})$, hence $\mu(\Gamma) = \mu(\rho)+kT\ln(K_{wall})$. The comparison shown in Figure 2 is excellent. The average deviation between predicted and simulated chemical potential is 0.06, which is well inside the expected systematic error.

### 3.2 Perturbation theory for elastic spheres with long-range attraction

Using the elastic disk equation of state, we can predict the equation of state of elastic spheres with a long range attraction by adding a negative second virial term, $\delta f_n = -b_2 y$. Just adding an attractive virial term is a first order perturbation to the purely repulsive system. When due to the attractive force the structure of the system changes considerably, perturbation theory will fail, even if higher order corrections are included. Therefore this approach may be used only for small perturbations and long-range interaction forces. To first order, the pressure and chemical potential become

$$\Pi = \Pi^{ED} - b_2 y \Gamma \quad ; \quad \mu = \mu^{ED} - 2b_2 y \qquad 15$$

where $\Pi^{ED}$ and $\mu^{ED}$ refer to the elastic disk results without long-range attraction.

Figure 3 shows the chemical potential for a number of values of $b_2$, for elastic spheres of repulsion $ad/kT$ = 1000. It is observed that for stronger attractions the chemical potential bends down as function of surface density. It has a horizontal inflection point for $b_2$ = 5.432. For larger values, liquid-vapour coexistence is predicted, where the coexistence densities follow from pressure and chemical potential equilibrium between both phases. The bimodal is indicated in Figure 3 by the black dashed curve.

Because the effective hard core diameter in the model changes when the particle elasticity is changed, the critical point also shifts if the particle elasticity is changed. The critical point is found for the $b_2$ value for which the first and second derivatives of the chemical potential vanish at the same density, the critical density. For strong repulsion, we find a deviation from hard core behaviour proportional to $(ad)^{-1/2} \propto E^{-1/2} d^{-3/2}$. Thus, the deviation from hard core behaviour is stronger for softer particles and particularly for small particles. A very good fit describing the critical point as function of the repulsion parameter is given by

$$b_2^{crit} \approx 5.914 - \frac{14.99}{\sqrt{ad/kT}}\left[1 - \exp(-\frac{\sqrt{ad/kT}}{1.85})\right] \qquad 16$$

The power ½ appearing in Eqs 12, 13 and 16, can be traced back to the force law in Eq 1. We may study a more general repulsive interaction force, which follows a power law:

$$F^{Rep} = a\left(1 - \frac{r}{d}\right)^{\mu-1} \qquad (r < d) \qquad 17$$

This includes the Hertz model,[30] for which $\mu = 5/2$. For a general power law repulsion, the second virial coefficient can still be calculated analytically in the limit of large repulsion. Its ratio to the hard disk second virial coefficient then follows

$$b = 2\int_0^1\left[1 - \exp\left(-ad(1-r)^\mu/\mu kT\right)\right]r dr$$
$$\approx 1 - \frac{2}{\mu}\left(\frac{\mu kT}{ad}\right)^{1/\mu}\Gamma(1/\mu) + \frac{2}{\mu}\left(\frac{\mu kT}{ad}\right)^{2/\mu}\Gamma(2/\mu) \qquad 18$$

where $\Gamma(x)$ is the standard Gamma function. This approximation is an asymptotic limit, the relative error is roughly $\mu \exp(-2ad/\mu kT)$. As a rule of thumb it is valid for $ad/kT > 5$. The error decreases rapidly as $a$ increases. Smaller repulsion parameters are not meaningful. In particular, for the Hertz model $\mu = 2.5$, and we obtain

$$b \approx 1 - 2.5601108\left(\frac{kT}{ad}\right)^{0.4} + 1.9385663\left(\frac{kT}{ad}\right)^{0.8} \qquad 19$$

Since we have previously found that $\lambda \propto b$ for $\mu = 2$, we may assume a similar proportionality for $\mu \neq 2$. The deviation of the spreading pressure from the hard core model is thus expected to be of order $(ad)^{-0.4} \sim E^{-0.4}d^{-1.2}$ for large repulsion. A full treatment would require a series of simulations to obtain the scaling parameter $\lambda$ as function of repulsion parameter $a$.

**3.3 Comparing theory and experiment**

To apply the theory to a particular experimental system we need to make one extra crucial step: mapping the physical particles onto simulation spheres. A correct description starts at the atomistic level and so includes all molecular subunits of a colloidal particle. Only if internal degrees of freedom are frozen out is it allowed to abstract a (gel) particle by an inert sphere. We identify two extreme cases, 1) all internal degrees of freedom in a particle are frozen, and 2) particles are polymers with many internal degrees of freedom. In case 1) each particle contributes $kT$ to the thermal energy, therefore we can identify one particle with one simulation sphere. The surface pressure at fixed area fraction then scales as $\Pi \sim kT/d^2$ where $d$ is the *particle* size. The proportionality is basically the compressibility factor. In case 2) we cannot neglect the internal degrees of freedom of the particles. At high area fraction ($y>0.2$) particles are frequently in contact so that internal degrees of freedom also contribute to the pressure. In this case the simulations provide an estimate for the compressibility factor, but the pressure essentially scales as $\Pi \sim kT/d_m^2$ where $d_m$ is the *monomer* size. Similar concepts are found in scaling theory of polymers,[31] where pressure is proportional to $\Pi \sim kT/d_m^3$ but independent of polymer length. The excluded volume interaction then serves to increase the proportionality between $\Pi$ and $kT/d_m^3$ because of the long-range correlation between monomers of the same polymer. We now discuss this on the basis of some examples.

An example approaching category 1) could be a system of glass beads, provided that long-range electrostatic repulsion and a short-range attraction are absent. Here we expect that many internal degrees of freedom are frozen. We calculate the spreading pressure for an area fraction $y = 50\%$. The dimensionless spreading pressure is $\Pi d^2/kT = 2.5$, hence the pressure in physical units is $\Pi = 2.5 \cdot kT/d^2$. For a particle diameter of $d = 100$ nm and $kT = 4.04 \cdot 10^{-21}$ J this gives a value of $10^{-3}$ mN/m. Experiments on glass beads[32] of 75 µm diameter give a pressure of 1 mN/m at a coverage of 75%, which implies $\Pi \approx 0.1$ mN/m at $y = 0.5$.

Examples of category 2) are polymer gel systems. Here a much higher surface pressure is found. An experimental value[33] for the spreading pressure of 113 nm size hyperbranched polyvinyl/polystyrene gel particles, at 50% area fraction, is 9 mN/m. Similar surface pressures are obtained for poly(lactic acid) particles at high pH, where the particles do not agglomerate.[34] The latter particles are 243 nm in diameter. Naïve application of the theory would predict $8 \cdot 10^{-4}$ mN/m in the former case and $2 \cdot 10^{-4}$ mN/m in the latter. Hence, if we were to make the identification that each polymer corresponds to one inert simulation particle, we would make an error by 4 to 5 orders of magnitude. To get a surface pressure of the right order of magnitude, we infer a length scale of 0.6 nm (see below). This implies that the density of monomers inside the particle is responsible for the pressure, and not just the interaction between particles. As stated above, the total number of degrees of freedom in the interface is responsible for the pressure, and not just the translational degrees. Each degree of freedom contributes to momentum transport, hence to surface pressure.

To find the number of degrees of freedom in the interface, we need to count the number of molecular units that are cut by the surface of stress. If we assume spherical particles and cut through one particle, the number of monomer units hit is given by $N_m = \pi \rho_m d^2 d_m/4$, where $\rho_m$ is the number density of molecular units in the particle and $d_m$ is the diameter of the molecular unit. The molecular density $\rho_m$ can now be expressed in the molecular volume fraction within a particle, $\phi_m = \pi \rho_m d_m^3/6$. Hence the number of molecular units in the interface that belong to the same particle is given by $N_m = \tfrac{3}{2}\phi_m(d/d_m)^2$. Since the number of particles per unit of area is given by $\Gamma$, the concentration of *molecular* units in the interface is given by

$$c = \Gamma N_m = \tfrac{3}{2}\phi_m \Gamma(d/d_m)^2 = \tfrac{6}{\pi}\phi_m y d_m^{-2} \qquad 20$$

To find the pressure, we simply count degrees of freedom in the interface. Each contributes $kT$ to the pressure. The ideal gas term adds $N_m kT$ per particle, but because of the entropy loss due to the connectivity between the monomers we have to subtract $(N_m-1)kT$. This leaves just one $kT$ per particle, leading to the ideal pressure $ckT/N_m$.

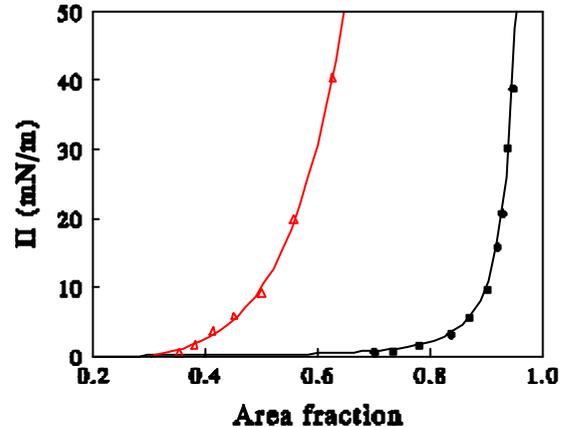

**Figure 4** Experimental data for glass beads (black dots) reproduced from Ref [32] and for gel particles (red triangles) reproduced from Ref [33], together with theoretical fits curves.

The interaction term is given by the density of monomers, multiplied by the excess compressibility factor $Z$. Since the compressibility factor is determined by the pair correlation function, which in turn is dominated by the structure *of the particles*, the compressibility factor of the system equals that of the particles. Thus spreading pressure of the polymers particles at the water-air interface is thus given by

$$\begin{aligned}
\Pi &= ckT/N_m + ckT\left(bZ(\lambda y)/\lambda - b_2 y\right) \\
&= \Gamma kT + N_m \Gamma kT\left(bZ(\lambda y)/\lambda - b_2 y\right) \qquad 21 \\
&\approx \tfrac{4}{\pi}\dfrac{kT}{d_{eff}^2}\left(byZ(\lambda y)/\lambda - b_2 y^2\right)
\end{aligned}$$

where $d_{eff}$ is the effective diameter of correlated domains within a particle. If all internal degrees of freedom within the particle are frozen $d_{eff} = d$, but in general $d_m < d_{eff} < d$.

We now return to two experiments mentioned above, the glass beads system studied by Hórvölgy et al,[32] and the gel particles studied by Wolert et al.[33] For both systems we assume $b = \lambda = 1$, and calculated the area fraction from the reported contact cross-sectional area. Pressure data was read off Figure 5a (curve A) of Ref [32], and from Figure 7 (113 nm curve) of Ref [33]. For the glass beads we make the fair assumption that $b_2 = 0$; for the gel particles we use $b_2$ as a fit parameter. A correction to the area fraction in both cases was less than 1%. The fit curves are shown in Figure 4. We find $d_{eff} = 7.1 \pm 0.4$ nm for the glass beads and $d_{eff} = 0.58 \pm 0.1$ nm for the gel particles, i.e. reasonable values for the correlated domains. The error includes uncertainty in the area fraction.

## 4. Elastic spheres with short-range attraction

Elastic spheres with a strong short-range interaction have been studied previously by Groot and Stoyanov.[12] They determined the phase diagram for a system at the critical density, as function of the interaction range and the temperature divided by the particle-particle adhesion energy. The phase diagram has been reproduced here schematically in Figure 5.

This diagram shows that for a short range interaction the thermodynamically stable state is quasi-crystalline at low temperature (high association energy G) or supercritical at high temperature (low association energy G). At a force range larger than $\delta = 0.2$ a liquid phase appears between solid and vapour. The location of a glassy state is merely indicative, it was observed at zero temperature.[12] Though not apparent on this scale, a liquid phase was predicted to reappear at very short interaction range $\delta < 10^{-3}$. A liquid phase for such a short force range has indeed been found recently in experiments.[35]

For typical hydrophobic particles of 100 nm diameter we estimate the interaction parameters as: $G/kT = 1000$ and $\delta =$

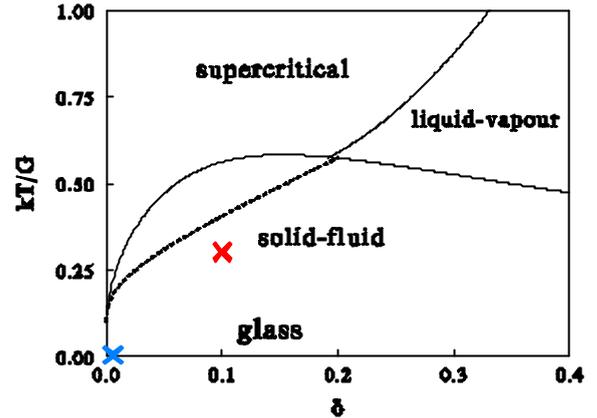

**Figure 5** Phase diagram of sticky elastic spheres in 3D at the (near critical) density $\rho d^3 = 0.512$, schematic after Ref [12]. The dashed curve is a line of hidden critical points. The red cross marks the position of the current simulations; the blue cross marks realistic parameters for 100 nm hydrophobic polymer particles.

0.007. This means that particle systems are at the bottom of the phase diagram in Figure 5, marked by the blue cross. For these parameters the stable state is quasi-crystalline. If we were to simulate this system with the adhesive hard sphere model, we would exclude the true equilibrium state. In contrast, the presently used sticky elastic sphere model does have a crystalline state.

In this section we will study the qualitative behaviour of adsorbing particles with a short-range attraction. To study this, it isn't necessary to reproduce all parameters, but if the stable phase is solid, then the simulated system should be in the solid phase. Thus, to reproduce the behaviour of uncharged (hydrophobic) particles, we fixed the repulsion parameter at $a = 2000$ and the temperature at $kT = 1$, and used maximum force $\varepsilon = 50$ and force range $\delta = 0.1$. These parameters correspond to adsorption energy $G/kT = {}^2/_3\ \varepsilon\delta d = 3.33$. This puts us deep enough into the solid phase to prevent evaporation of clusters, but it allows much faster simulation than for the real adsorption energies and force range.

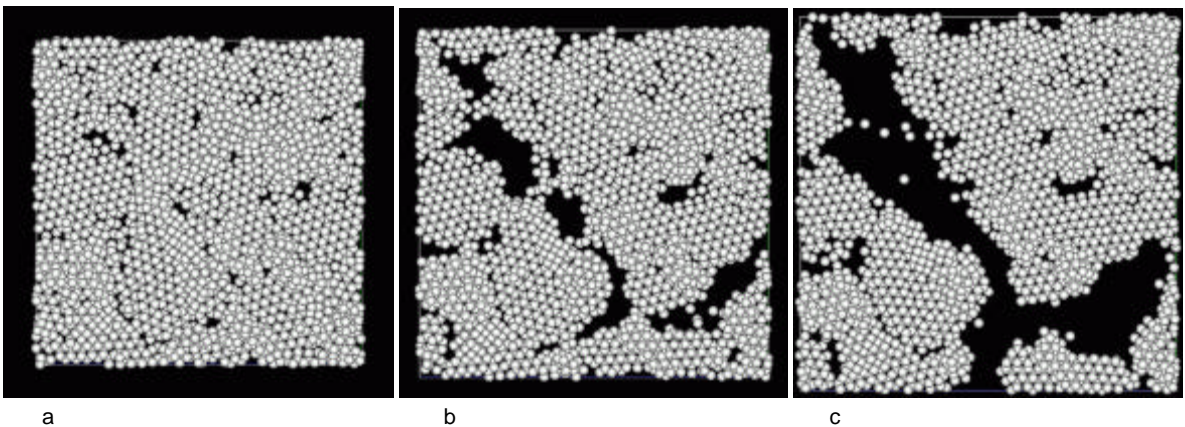

**Figure 6** Conformation of sample without deformation (a), for an area increase by 16% (b), and by 35% (c).

In the first system studied, we used a box of sizes 3×30×30 and inserted 2700 particles randomly throughout the box. The friction factor was initially put at $\gamma = 100$; the step size was set at $\delta t = 0.005$, and the wall adsorption constant was chosen as $K_{wall} = 1000$. After the initial particle insertion, temperature starts at a high value because particles overlap, but after about 100 to 200 steps the system has cooled down to $kT = 1$. Note that the density in the system is $\rho d^3 = 1$, which is a typical liquid density. At this point in time the system size in the $x$-direction was enlarged to $L_x = 10$, and we switched to lower friction $\gamma = 10$. This gives the system the opportunity to build up liquid-vapour coexistence, with an adsorbed liquid layer to the left wall. For the parameters chosen, however, the liquid phase is unstable relative to the solid phase (see Figure 5) and indeed crystalline domains appear. Because the density in a crystal phase is higher than in the liquid phase at the same pressure, the solid contracts and a number of small holes between crystalline domains have appeared at time $t = 100$. This conformation is shown in Figure 6a. To study the system behaviour under deformation, the simulation box was expanded in the $y$ and $z$-direction. The area was expanded by 0.06% every 20 time steps, over a total of $10^4$ steps. Thus, the area was increased by 35% over the whole run. The conformations at the beginning, at half time and at the end of the run are shown in Figure 6. This shows that the system fails through fault lines and then moves as "drifting continents".

Next we explore the influence of the particle concentration by changing the box size in $y$ and $z$-directions to 39×39, 50×50, and 70×70. The same number of particles and the same box size in the $x$-direction were taken as above. We thus start with the overall densities $\rho = 0.59$, $\rho = 0.36$ and $\rho = 0.18$. These systems were again prepared using friction $\gamma = 100$ and evolved until the temperature and pressure had equilibrated. At time $t = 1$, the friction factor was reduced to $\gamma = 10$, and the system evolution was followed.

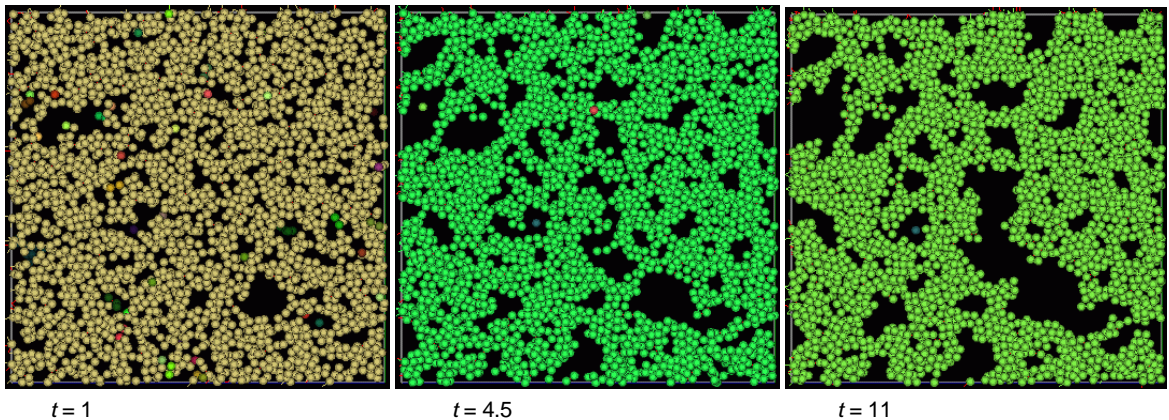

$t = 1$      $t = 4.5$      $t = 11$

**Figure 7** Conformation at density $\rho = 0.59$ evolving over time. Colours indicate connected clusters.

At density $\rho = 0.59$, no homogeneous liquid film is formed. Instead, we have a phase separating liquid, because the temperature ($kT/G = 0.3$, red cross in Figure 5) is below the (hidden) critical point[12] ($kT/G^c \approx 0.405$) and the density is below the liquid density. In fact the density is close to the critical density, so we may expect a spinodal phase separation process. Indeed, the pattern formed is characteristic of a spinodal process, where irregular structures are formed at one dominant length scale, see Figure 7. Since the liquid phase is actually unstable relative to the crystal phase (see Figure 5) it must eventually freeze. This happens around $t = 11$, which halts the phase separation process. Therefore, the structure seen at $t = 50$ is virtually the same as at $t = 11$. Note that the colours in Figure 7 indicate connected clusters. This shows that from the first moment on, virtually all particles are connected in the same percolating cluster.

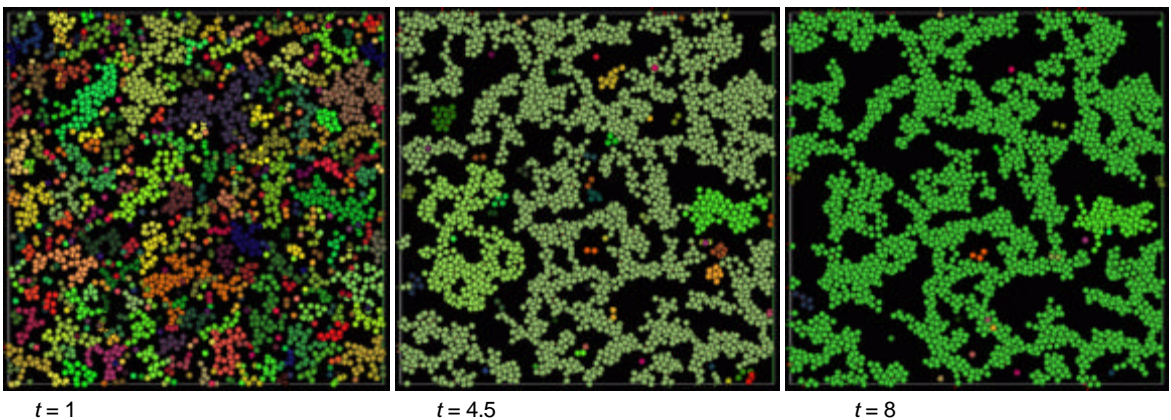

$t = 1$      $t = 4.5$      $t = 8$

**Figure 8** Conformation at density $\rho = 0.36$ evolving over time. Colours indicate connected clusters.

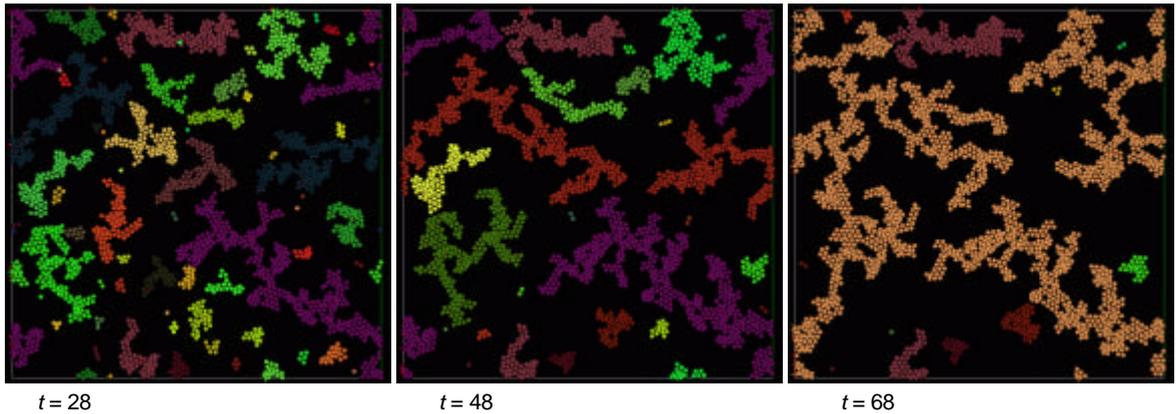

| $t = 28$ | $t = 48$ | $t = 68$ |

**Figure 9** Conformation at density $\rho = 0.18$ evolving over time. Colours indicate connected clusters.

At the lower density $\rho = 0.36$ the system was again equilibrated with high friction until time $t = 1$, and then it was evolved to $t = 8$. In sharp contrast with the above system, the system at $t = 1$ consists of many disconnected clusters, see Figure 8. These clusters grow together, and a percolating cluster is formed around $t = 3$. At that point in time, the system forms a surface-adsorbed fractal-like structure, see Figure 8. This simulation mimics the formation of particle layers at the water-air surface, when the particles are slowly added to water. In that case the particles are expected to agglomerate at the surface, forming a surface fractal when they meet other particles. In the simulated system the clusters are first formed in the bulk, then they adsorb to the surface, and form a network.

When the system is subsequently stretched, the fractal structure is expanded and bundles of particles are observed to move over the surface. These bundles stick together at junction points, where the structure reorganizes like a liquid. Between the junction points the particles are ordered in a crystalline structure. In fact the system at $t = 8$ is not yet in a stationary state. If we evolve the original system further, it keeps on coarsening until about $t = 50$, where it freezes. If we reduce the density further to $\rho = 0.18$, many particles adsorb at the surface before they associate to each other in the bulk. Therefore, two-dimensional clusters appear at the surface with only few particles on top. Figure 9 shows the evolution in time. This structure is clearly formed by cluster-cluster aggregation, and therefore it must be fractal. We note the striking similarity between our simulation results of particles with a short-range attraction in Figures 8 and 9 and the observations by Reynaert et al.[36] on polystyrene particles on water-air and water-decane interfaces.

At this lowest density ($\rho = 0.18$) the time to form a percolating cluster is much larger than for $\rho = 0.36$ (see Figure 8). Moreover, in both cases ($\rho = 0.18$ and $\rho = 0.36$) only a single path is formed that spans the network from left to right and from top to bottom when all particles are connected. This indicates that both networks are very weak, because the modulus is proportional to the density of stress bearing network connections. It also indicates that the systems shown here are too small to determine the modulus directly at low density.

Therefore we need to use an indirect method to predict the modulus of a large colloidal particle network in 2D. In three dimensions, scaling relations for the modulus as function of volume fraction are given by Kantor and Webman,[37] and by Jones and Marques.[38] Simulations and scaling relations were given by West et al.[39] Using the reasoning by Jones and Marques, a network of cross-link functionality $n$, made of stiff connections of N segments, has a volume fraction $y = nN/R^d$, where $R$ is the size of the network connection and $d$ is the space dimension; hence $y/N = n/R^d$. For a network where the modulus is dominated by the bending rigidity $B$ of the segments, the modulus thus follows as[38] $G' = Bn/NR^d = By/N^2$. If the particle clusters are characterised by $N \sim R^{1/\nu}$, where $1/\nu = d_f$ is the fractal dimension, we find the modulus as

$$G' \sim E \sim \Pi \sim y^{1-2/(1-d\mathbf{n})} \sim y^{(2+d_f)/(2-d_f)} \qquad \mathbf{22}$$

In the last step $d = 2$ was substituted for the space dimension. Within a simple scaling approximation the osmotic pressure should behave in the same way as the shear modulus.

To make a general prediction, we need to know the fractal dimension of the network. For cluster-cluster aggregation, as we have here, this depends on whether we have diffusion limited aggregation (DLCA) or reaction limited aggregation (RLCA). Simulation predictions were given first by Meakin,[40] and thorough experimental studies were done by Robinson and Earnshaw.[41] They report values for the fractal dimension as $d_f = 1.44 \pm 0.04$ for DLCA in two dimensions, and $d_f = 1.55 \pm 0.03$ for RLCA. The first case occurs when sticking interaction between particles is fast, and the second is pertinent when particles have to collide many times before they stick. However, the picture is more complex, as fractal dimensions as high as $d_f = 1.75 \pm 0.07$ are reported in experiments (see Ref [41] and references therein) for 2D systems at high initial density, close to the sol-gel transition. This is in line with theory and simulations.[42,43] Another extreme is found when electrostatic polarisation of adsorbed clusters is considered, which may produce fractal dimensions as low as $d_f = 1.26 \pm 0.06$.

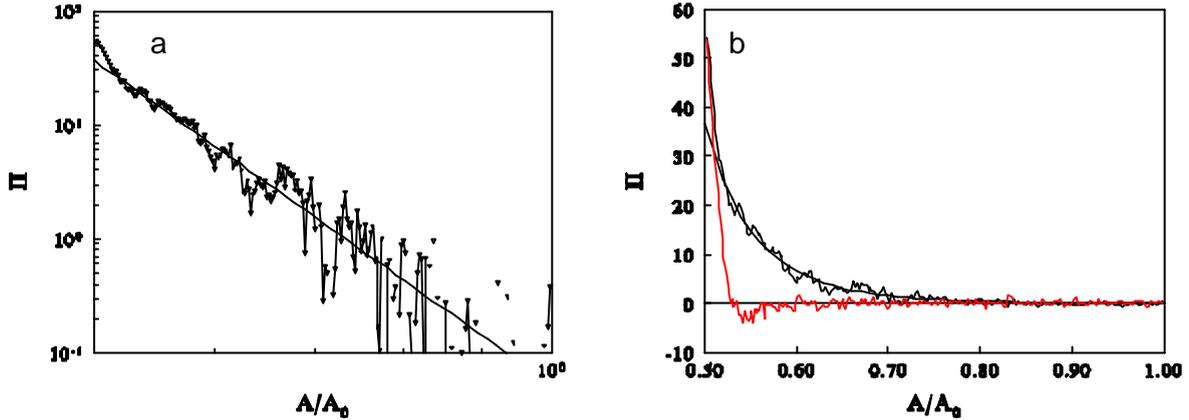

**Figure 10 a** (left): Spreading pressure of ρ = 0.36 system under compression. The line has slope –9.5. **b** (right) Spreading pressure during compression (black) and subsequent expansion (red curve).

The fractal dimension of a network structure is furthermore reported to depend on concentration. For concentrations at and above the percolation concentration, the entire spanning cluster in a network scales as the percolating cluster obtained by standard percolation.[44] The fractal dimension of a percolating cluster in 2D is given by[45] $d_f = 1.62\pm0.02$, but for simulated networks Hasmy and Jullien[44] report a fractal dimension of the backbone that increases continuously with concentration up to $d_f = 2$. This would imply that there is no simple scaling law of modulus with concentration, even though the clusters show a power law pair correlation at any given concentration. Thus, as concentration increases one may expect the modulus roughly to behave as

$$G' \sim \Pi \sim \Gamma^{6.1\pm0.5} \quad ... \quad G' \sim \Pi \sim \Gamma^{9.5\pm0.5} \qquad 23$$

The former power law is expected for *diffusion-limited cluster-cluster aggregation* at low dilution. The latter power law is based on 2D *percolation*, hence is expected to be valid close to the sol-gel transition. As the concentration is further increased, the modulus can be expected to diverge when the fractal dimension reaches 2 and the solid particles are pushed into each other. To test this fractal model, the system of density ρ = 0.36 was compressed by gradually reducing the surface area. The area was reduced by 0.06% after every 20 time steps. Since the time step taken is $\delta t = 0.005$, the strain rate in simulation units is $d\ln(A)/dt = -0.006$. The resulting compression curve is shown in Figure 10a. The slope in a log-log plot to the area indeed equals –9.5±0.5. Within the error this is in line with the prediction in Eq 23 that is based on *percolation exponents*.

Under expansion the system shows a completely different behaviour. The spreading pressure now declines sharply, it becomes negative, and then the system breaks, see red curve in Figure 10b. During compression the fractal structure of the system was destroyed irreversibly, as it was compressed to a 2D structure with a few imperfections. On subsequent stretching it breaks at these imperfections and the system expands like drifting continents, as described before for a system that was prepared at high density.

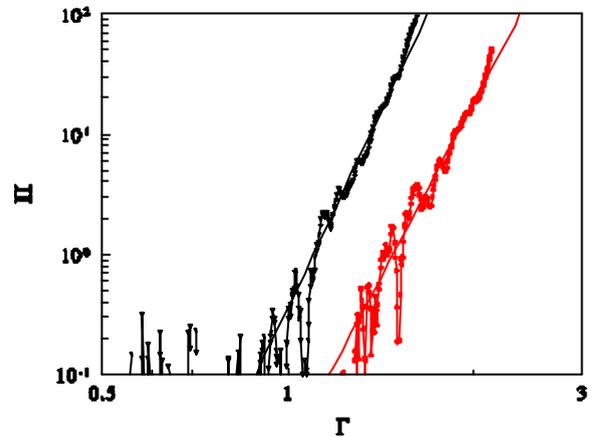

**Figure 11** Spreading pressure during compression for density ρ = 0.18 (black curve) and ρ = 0.36 (red curve). The slopes are equal within the error, but the low density system has a higher spreading pressure at the same surface density.

If we compress the ρ = 0.18 system in the same way, a pressure curve is found that is shifted *upwards* because the clusters at the surface are thinner. This leads to a higher effective surface density. The slope in a log-log plot to the area fraction is slightly higher than for ρ = 0.36. We find a slope 10.7±0.5, see Figure 11. Within the accuracy of the simulations this is the same as for the ρ = 0.18 system. Just as the structures formed in these simulations, the strength of the networks corresponds well to the experimental value obtained by Reynaert *et al*,[36] who find the exponent 13±2.

To summarise the observations, if we prepare our system within a narrow layer from the surface at low density, particles will adsorb at the surface and form a surface-adsorbed fractal structure. Small clusters can either form in the bulk and subsequently deposit, or colloidal particles may stick to a cluster that was previously adsorbed. In either case, the surface-adsorbed clusters will stick together when they meet, leading to a fractal structure formed by cluster-cluster aggregation. At intermediate density the colloidal particles form an instable liquid layer in which holes grow by a spinodal process. This process is halted when the (unstable) liquids freezes. Finally, at high density no phase-separation occurs, and the liquid phase freezes into a polycrystalline solid. When a system is adsorbed at low surface density, it forms a surface fractal network. Compression of such a system gives a high power law for the spreading pressure as function of surface coverage, roughly $\Pi \sim \Gamma^{10}$. The power is apparently determined by 2D percolation. Compression changes the structure irreversibly. Expansion of a once compressed system gives a very rapid decline in pressure, ultimately leading to rupture of the structure.

## 5. Summary and conclusions

Foam and emulsion stabilization by particles that adsorb on the air-water interface is linked to colloidal stability. When the particles have an attractive interaction, they form a solid network at the surface that provides surface elasticity. This elasticity prevents disproportionation of bubbles, i.e. swelling of large bubbles to the expense of small ones.

To study this theoretically, a simulation model has been developed for particles that can adsorb to and desorb from a water-air interface. First, an equation of state for purely repulsive elastic spheres is developed. Based on this, a perturbation theory for elastic spheres with a long-range attractive interaction is formulated.

The present simulations show that without attractive interaction, particles would still form a solid at high surface coverage. However, such a solid cannot withstand negative pressure. This indicates that forming a solid structure is not sufficient to provide foam stability. We speculate that foam stability may be related to the ability to withstand negative pressures, or to the bending rigidity of the surfaces.

A new insight arises from comparing simulation to experiment: the surface pressure is generally *not* determined by the pressure of the particles alone, interacting via some effective potential. Internal degrees of freedom *within* the particles are all-important. Consequently, the surface pressure of a fully packed layer is not proportional to $kT/d^2$, where $d$ is the particle diameter; instead for gel particles it is proportional to $kT/d_m^2$, where $d_m$ is the size of the *molecular* building blocks of the gel polymers. For a glass bead system we conclude that the surface pressure is proportional to $kT/d_{eff}^2$, where $d_{eff}$ ($> d_m$) corresponds to the size of correlated domains within the particles.

Finally, simulations were done for elastic spheres with a short-range attraction. At low surface coverage these show the occurrence of fractal-like structures. At intermediate surface coverage we find irregular structures that resemble a spinodal system. However, at realistic interaction strength, the colloidal system is expected to freeze, and indeed the spinodal structure was arrested. At high surface coverage the simulations show poly-crystalline domains.

Simulation of compression of surface layers formed in the dilute limit predicts power laws for the pressure as function of area. This power law has been related to the (fractal) structure of the adsorbed network.

**Acknowledgments:** The authors wish to thank Dr. Theo Blijdenstein and Mr. Peter de Groot from Unilever Research in Vlaardingen, The Netherlands for numerous stimulating discussions.

## Notes and references

‡ To test the method to generate an adsorption equilibrium, 800 ideal gas particles were enclosed in a box of size 5x28x28, of which half (A) were adsorbing to the left wall with $K = 5$, and the other (B) half were non-adsorbing. If the adsorbed particles have a vanishing velocity *perpendicular* to the surface we find clustering of the A and B particles at the wall and an increasing concentration of B particles towards the wall. The reason is that a vanishing perpendicular velocity violates the fluctuation-dissipation balance of the thermostat – the adsorbed particles are cold. Therefore, adsorbed particles should be given a random perpendicular velocity. We choose these from the uniform distribution $(-\sqrt{3}, \sqrt{3})$, which produces the correct variance. The resulting density profile for the B particles is then uniform, but the density profile of the A particles remains slightly inhomogeneous.

To track down the cause of this inhomogeneity, adsorption at a lower exchange rate was studied. We used a noise and friction cut-off range $r_c = 2$, and simulated over $4 \times 10^5$ time steps of $\delta t = 0.06$ with friction factor $\gamma = 3$. If we simulate a high adsorption constant $K_{wall}$ and put $\omega = 1$, every collision would result in adsorption (see Eq 7 and the paragraph above). This is diffusion-limited adsorption. The consequence is that particles have a high probability of re-adsorbing immediately after desorption. Therefore the density profile has no time to equilibrate. The excess adsorption over reaction-limited adsorption is proportional to the adsorbed amount; hence the adsorption equilibrium for the ideal gas takes the form $\Gamma = K_{wall}\rho + \alpha\Gamma$. Empirically we find the excess adsorption to be proportional to $\sqrt{\omega}$. For the simulation parameters used we find $\alpha = (0.177\pm0.004) \sqrt{\omega}$, which rescales $K_{wall}$ by a factor $1/(1-0.18\sqrt{\omega})$. Truly reaction-limited adsorption can thus be obtained by simulating at two values of $\omega$ and extrapolating to $\omega = 0$. For diffusion-limited adsorption we simply put $\omega = 1$. For surfaces with adsorbed (hard) particles the adsorption process to a filled surface is by nature reaction limited. In that case we may safely put $\omega = 1$ to obtain the true equilibrium.